\documentclass[aip,amsmath,amssymb,reprint]{revtex4-1}
\usepackage[english]{babel}
\usepackage{graphicx}
\usepackage{dcolumn}
\usepackage{bm}
\usepackage{epstopdf}
\usepackage[utf8]{inputenc}
\usepackage[T1]{fontenc}
\usepackage{mathptmx}
\usepackage{natbib}
\usepackage{color}
\renewcommand{\Re}{\mathrm{Re}}

\begin{document}
\preprint{AIP/123-QED}
\newcommand{\JH}[1]{\textcolor{blue}{JH: #1}}
\newcommand{\TN}[1]{\textcolor{magenta}{TN: #1}}

\title[Inertial migration of oblate spheroids in a plane channel]{Inertial migration of oblate spheroids in a plane channel}
\author{Tatiana V. Nizkaya}
\affiliation{Frumkin Institute of Physical Chemistry and
Electrochemistry, Russian Academy of Science,\\ 31 Leninsky Prospect,
119071 Moscow, Russia}%

\author{Anna S. Gekova}
\affiliation{Frumkin Institute of Physical Chemistry and
Electrochemistry, Russian Academy of Science,\\ 31 Leninsky Prospect,
119071 Moscow, Russia}

\author{Jens Harting}
\affiliation{Helmholtz Institute Erlangen-N\"urnberg for Renewable Energy, Forschungszentrum J\"ulich,\\ F\"urther Str. 248, 90429 N\"{u}rnberg, Germany}
\affiliation{Department of Chemical and Biological Engineering and Department of Physics,\\ Friedrich-Alexander-Universit\"at Erlangen-N\"urnberg, F\"urther Str. 248, 90429 N\"{u}rnberg, Germany}

 \author{Evgeny S. Asmolov}
\affiliation{Frumkin Institute of Physical Chemistry and
Electrochemistry, Russian Academy of Science,\\ 31 Leninsky Prospect,
119071 Moscow, Russia}
\author{Olga I. Vinogradova}
\email[Corresponding author: ]{oivinograd@yahoo.com}
\affiliation{Frumkin Institute of Physical Chemistry and
   Electrochemistry, Russian Academy of Science,\\ 31 Leninsky Prospect,
   119071 Moscow, Russia}

\date{\today}

\begin{abstract}
We discuss an inertial migration of oblate spheroids
in a plane channel, where steady laminar flow is generated by a pressure gradient.
Our lattice Boltzmann
simulations show that spheroids orient in the flow, so that their minor axis coincides with the vorticity direction (a log-rolling motion). Interestingly, for spheroids of moderate aspect ratios, the
equilibrium positions relative to the channel walls depend only on their equatorial radius $a$. By analysing the inertial lift force  we argue that this force is proportional to $a^3b$, where $b$ is the polar radius, and conclude that the dimensionless lift coefficient of the oblate spheroid does not depend on $b$, and is equal to that of  the sphere of radius $a$.

\end{abstract}

\maketitle

\section{Introduction}

It is well-known that at finite Reynolds numbers particles migrate across streamlines of the flow to some equilibrium positions in the microchannel. This migration is attributed to inertial lift forces, which are currently successfully used in microfluidic systems to focus and separate particles of
different sizes continuously, which is important for a wide range of applications~\cite{zhang2016,dicarlo19}. The rapid development
of an inertial microfluidics has raised a considerable interest in the lift forces on
particles in confined flows.
The majority of previous
work on lift forces has assumed that particles are spherical.
In their pioneering
experiments, Segr\`e and Silberberg found that small spheres focus to a narrow
annulus at a radial position of about 0.6 of a pipe radius\cite{Segre:Silb62a}.
Later, several theoretical\cite{Vas:Cox77,Asmolov99,hood2015,asmolov2018} and
numerical\cite{dicarlo2009prl} studies proposed useful scaling and
approximate expressions for the lift force in a channel flow, which are frequently invoked.  The assumption that particles are spherical often becomes unrealistic.
The non-sphericity could strongly modify the lift forces, so the shape of particles becomes a very important consideration~\cite{behdani2018}. The body of theoretical and experimental work
investigating lift forces on non-spherical particles
is much less than that for spheres, although
there is a growing literature in this area.

\citet{hur2011} and \citet{masaeli2012} appear to have been the first to study experimentally the inertial focusing of non-spherical
particles. These authors addressed themselves to the case of particles (spheres and rods of different aspect
ratios) of equal volume, and demonstrated the possibility of their separation in a planar channel of moderate Reynolds numbers, $\Re\leq
100$. \citet{roth2018} recently reported the separation of spheres, ellipsoids and peanut-shaped
particles in a spiral microfluidic device, where
the inertial lift force is balanced by the Dean force that can be generated in curved channels~\cite{dicarlo2007}. These papers concluded that a key parameter defining equilibrium positions of particles is their rotational diameter. The authors, however, could not relate their results neither to the variation of the lift force across the channel nor with its dependence on particle shape since these are unaccessible in experiment.

The theoretical analysis of the lift on non-spherical particles is beset with difficulties since they could  vary their orientation due to a rotation in a shear flow, which, in turn, could induce unsteady flow disturbances leading to a time-dependent lift~\cite{su2018}. There have been some attempts to provide a theory of such a motion by employing spheroids as a simplest model for non-spherical particles. It is known that at
vanishing particle Reynolds numbers, $\mathrm{Re_p}$, non-inertial spheroids exhibit in a shear flow a periodic kayaking motion  along one of the Jeffrey orbits~\cite{jeffrey1922}. However, the orientation of oblate spheroids of finite
$\mathrm{Re_p}$ eventually tends to a stable state due to inertia of the fluid and the particle~\cite{saffman1956}.

Computer simulations might shed some light on these phenomena, and, indeed, computational inertial microfluidics is  a growing field that currently attracts much research efforts~\cite{bazaz2020computational}.
There is a number of simulations using the lattice Boltzmann method (LBM) that is well-suited for parallel processing and allows one an efficient tracking of the particle-fluid interface~\cite{LaddVerberg2001}, which are directly relevant. A large fraction of these deal with
rotation properties of spheroids in shear flows~\cite{qi2003,janoschek11,rosen2014,huang2017}. At moderate $\mathrm{Re_p}$, oblate spheroids exhibit a log-rolling motion about their
minor axis oriented along the vorticity direction, while prolate particles tumble, but when $\mathrm{Re_p}> 200$, in some situations a transition to other rotational regimes may occur~\cite{qi2003}. Its  threshold depends on the particle aspect ratio, which can be used for their separation~\cite{li2017shape}.
 However, neither this paper addresses itself to the issues
of inertial migration. This was taken up only recently in the paper by \citet{lashgari2017}, who carried out simulations of stable
equilibrium positions and orientations of oblate spheroids  in rectangular channels. The lift
force on cylindrical particles in rectangular ducts has been calculated by \citet{su2018}. These authors found that particles execute a
periodic tumbling motion, so that the lift force is unsteady, but its
average dependence on the particle position, however, is similar to that for a sphere. Finally we should mention that\citet{huang2019} used
dissipative particle dynamics simulations to find the equilibrium positions for prolate and oblate spheroids in a plane
Poiseuille flow.

Nevertheless, in spite of its importance for separations of particles, the connection between the shape of non-spherical
particles and emerging lift forces to remain poorly understood. In this paper we present some results of an LBM study of the inertial migration of oblate spheroids in a plane channel, where steady laminar flow of moderate $\Re$ is generated by a pressure gradient. We perform measurements of the lift force acting on spheroids in the stable log-rolling regime and find that the lift coefficient depends only on their equatorial radius $a$. To interpret this result we develop a scaling theory and derive an expression for a lift force.
Our scaling expression has a power to easily predict equilibrium
positions of oblate spheroids in microfluidic channels.

Our paper is arranged as follows. In Sec.~\ref{sec:setup} we define our system and mention briefly some expressions for a lift force acting on a spherical particle. Sec.~\ref{sec:simulation} describes details and parameters of simulations. Simulation results are discussed in Sec.~\ref{sec:results}. We then present scaling arguments leading to an expression for a lift force.
We conclude in Sec~\ref{sec:conclusion}.

\section{Model}
\label{sec:setup}

\begin{figure}
\centering
\includegraphics[width=0.9\columnwidth]{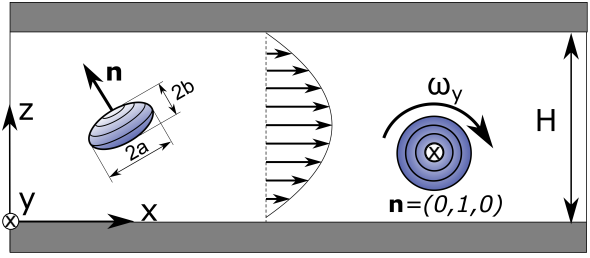}
    \caption{An oblate  spheroid  orienting  in  a    pressure-driven flow to perform a stable log-rolling state.}
  \label{fig:sketch}
\end{figure}

We consider an oblate spheroid with an equatorial radius $a$ and a polar radius
$b<a$ in a pressure-driven flow between two parallel walls, separated by a
distance $H$ (see Fig.~\ref{fig:sketch}). Its location in the channel is defined by coordinates of the center $\mathbf{x}=(x,y,z)$ and by a
unit vector directed along the symmetry axis $\mathbf{n}=(n_x,n_y,n_z)$ (referred below to as the orientation
vector). At the channel walls and particle surface we apply no-slip boundary
conditions.

The velocity profile in the channel in the absence of the particle is
parabolic,
\begin{equation}
	U(z)=4U_mz(1-z/H)/H,
\end{equation}
where $U_m=|\nabla p|H^2/(8\mu)$ is the
fluid velocity at the channel center, $\nabla p$ is a pressure gradient and
$\mu$ is the dynamic viscosity. The (finite) channel Reynolds number $\Re=\rho
U_mH/\mu$, where $\rho $ is the fluid density.

The inertial lift force drives
particles across the flow streamlines. For spherical particles it can be written as~\cite{Asmolov99,asmolov2018}
\begin{equation}
F_l(z)=\rho a^4 G_m^2 c_l, \label{lift}
\end{equation}
where $G_m=4U_m/H$ is the shear rate at the wall and $c_l$ is the lift
coefficient, given by
\begin{equation}
c_l=c_{l0}+V_s c_{l1}+V^2_s c_{l2}, \label{cli}
\end{equation}
where $c_{li}$, $i=0,1,2$ are the lift coefficients that depend on the dimensionless particle position $z/H$, its size $a/H$, and $\Re $. The dimensionless slip velocity is defined by
\begin{equation}
V_s = \dfrac{V^x_p-U(z)}{U_m},
\label{Vslip}
\end{equation}
where $V^x_p$ is the
$x$-component of the particle velocity and $U$ is the undisturbed fluid velocity at the
particle center $z$. Note that it is normally considered that the slip velocity is induced by external forces only and, consequently, does not impact a hydrodynamic lift of neutrally buoyant particles. However, it has been recently shown that for neutrally buoyant particles $V_s$ is negligibly small only in the central portion of  the channel, but not near the wall, where it becomes finite~\cite{asmolov2018}.

Eq.(\ref{lift}) is widely invoked to estimate the migration velocity
of neutrally buoyant (of $\rho_p = \rho$) spherical particles~\cite{dicarlo2007}. When the lift force is balanced by the Dean~\cite{dicarlo2007} or external~\cite{zhang2014real,dutz2017fractionation} force $F_{ex}$ (in the case of non-neutrally buoyant particles, $\rho_p \neq \rho$), Eq.(\ref{lift}) can be applied to find the equilibrium positions, $z_{eq}$, using the force balance
\begin{equation}\label{eq:balance}
 F_l(z_{eq})+F_{ex}=0
\end{equation}
One normally assumes that $F_{ex} = V f_{ex}$, where $V=\frac{4}{3}\pi a^{3}$ is the volume of a sphere and $f_{ex}$ is a force per unit volume. For instance, under the influence of gravity $f_{ex}=-\left( \rho _{p}-\rho \right) g$.
In order to employ a similar approach
to the shape-based separation of spheroids (of $V=\frac{4}{3}\pi a^{2}b$) it is necessary to know how the lift force scales
with the particle radii $a$ and $b$, and with the aspect ratio $b/a$.

It is of considerable interest to obtain a similar scaling equation for spheroids. However, as described in the Introduction, their instantaneous orientation and rotation are often functions of time, which should lead to a time-dependent lift. Nevertheless, for neutrally-buoyant
oblate spheroids of finite $\mathrm{Re_p}=\rho G_ma^2/\mu$, the symmetry axis eventually becomes parallel to
the vorticity direction, $\mathbf n_{eq}=(0,1,0)$ (the log-rolling motion). Consequently, to predict their long-term migration we have to find a lift force for this steady
configuration. Once it is known, the equilibrium
positions of oblate spheroids (including non-neutrally buoyant too) can be found by balancing the lift and external forces.

\section{Simulation setup}\label{sec:simulation}

To simulate fluid flow in the channel we use  a 3D, 19 velocity, single
relaxation time implementation of the lattice Boltzmann method (LBM) with a
Batnagar Gross Krook (BGK) collision
operator~\citep{benzi_lattice_1992,kunert2010random}. Spheroids are
discretized on the fluid lattice and implemented as moving no-slip boundaries
following the pioneering work of Ladd~\cite{LaddVerberg2001}.   Details of our
implementation can be found in our previous
publications\cite{janoschek2010b,kunert2010random,janoschek2014,Dubov14,asmolov2018,nizkaya2020}.

The size of the simulation domain is $(N_{x},N_{y},N_z)=(128,128,81)$, with
corresponding channel height $H=80$ (all units are simulation units).
No-slip boundaries are implemented at the top and bottom channel walls using
mid-grid bounce-back boundary conditions and all remaining boundaries are
periodic. The kinematic viscosity is $\nu=1/6$ and the fluid is initialized
with a density $\rho=1$. A body force directed along $x$ with volumetric
density $g=0.5\dots 2 \times10^{-5}$ is applied both to the fluid and the
particle, resulting in a Poiseuille flow with $\Re \simeq 11...44$.

To simulate particle trajectories we use spheroids of equatorial radii $a=6$, $8$ and $12$
and several aspect ratios, $0.33\leq b/a\leq 1$. This range of particle aspect ratios is chosen to ensure the correct representation of ellipsoidal shape on the grid.
The particles start close to the expected
equilibrium with zero initial velocity and in log-rolling orientation.
We assume that the equilibrium  is reached when the difference between an average of particle $z$-coordinate over 10 time steps and its average over the next 10 step does not exceed $1.25\times10^{-6}H$.

To measure the lift force as a function of $z$, we fix particle $z$-coordinate but let to rotate and to move in all other directions. Particle motion starts with
zero initial velocity and $\mathbf{n}=(0,1,0)$ that corresponds to the stable
log-rolling state. Once a stationary velocity is reached, the vertical
component of the force $F_l(z)$ is measured and is averaged over $10^4$ simulation steps. Therefore, these measurements also correspond (if we neglect force fluctuations) to non-neutrally particles at equilibrium, Eq.\eqref{eq:balance}.

To check if the results depend on the box size due to periodic boundary
conditions in $x-$ and $y-$directions, we also simulate the migration of the
large sphere of $a=b=12$ in a larger simulation box with
$(N_{x},N_{y},N_z)=(256,256,81)$. The difference in equilibrium positions for
the two box sizes is $100$ times smaller than the typical separation of
equilibrium positions of different particles.

\begin{figure}
\vspace{-0.4cm}
\centering
\includegraphics[width=1.0\columnwidth]{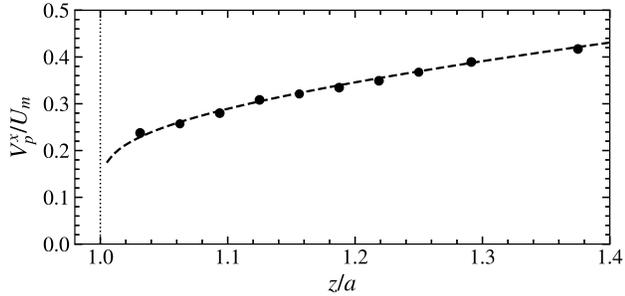}
\vspace{-1.2cm}
\caption{Velocity of a  sphere of $a/H=0.1$ located at a distance $z$ from the wall and free to rotate and translate in the $x$-direction (circles). Dashed curve are calculations from Eq.~\eqref{gold} representing the semi-analytical solution for a wall-bounded shear flow. Dotted line indicates a contact with the wall. }
\label{fig:near_wall}
\end{figure}

To test the resolution of the method in the near-wall zone, we measure the velocity of the freely rotating and translating in $x$-direction sphere of radius $a=8$, which $z$-coordinate is fixed. The $x$-component of the velocity $V_p^x$ is plotted in Fig.~\ref{fig:near_wall}, along with a semi-analytical solution for a wall-bounded shear flow \cite{reschiglian2000}(see Eq. \eqref{gold}). One can see that a sufficient accuracy is attained for separations as small as 1 lattice nodes ($z/a>1.05$), similarly to previous results for the sphere approaching a rough wall~\cite{kunert2010random}.

\section{Numerical results and discussion}\label{sec:results}


\begin{figure}
\vspace{-0.4cm}
\centering
\includegraphics[width=1.0\columnwidth]{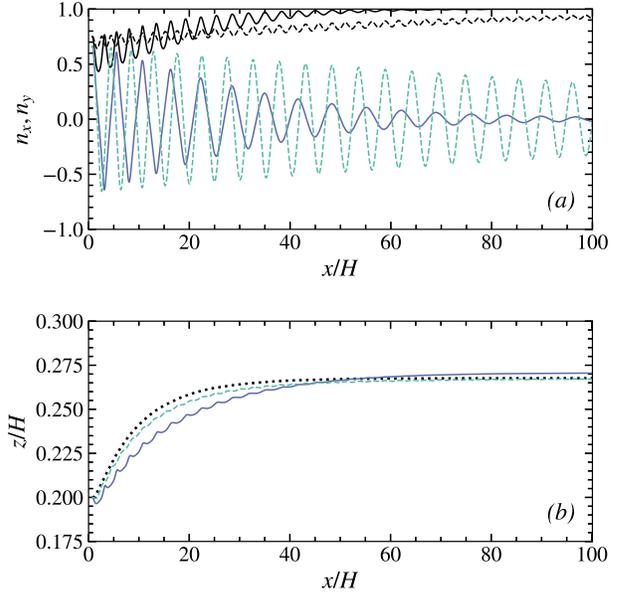}
\vspace{-1.4cm}
\caption{(a) $x$-components (colored curves) and $y$-components (black curves) of the orientation vector and (b) trajectories for spheroids with $a/H=0.15$ and $b/a=0.5$ (solid), $0.8$ (dashed) and $1$ (dotted).
	}
\label{fig:traj}
\end{figure}

We first simulate trajectories and orientations of freely moving neutrally buoyant
spheroids of different sizes and aspect ratios in a flow with $\Re=22$.
Our results show that regardless of the initial position and orientation, oblate spheroids eventually
reorient to the stable log-rolling motion around the axis of symmetry and their
angular velocity is directed along  the $y$ axis, $\mathbf{n}=(0,1,0)$,
$\boldsymbol{\omega}=(0,\omega_y,0)$. We also observe that they focus at some distance $z_{eq}$
from the wall due to the inertial migration. The rates of reorientation and
migration depend on the particle size and the aspect ratio.

In Fig.~\ref{fig:traj} we compare the rotational behavior and trajectories of
spheroids of several aspect ratios, $b/a=1$ (sphere), $0.8$ and $0.5$, but of the
same equatorial radius $a/H=0.15$. For all simulations the initial position and orientation
 are fixed to $z_0/H=0.2$ and $\mathbf{n}_0=(0.66,0.75,0)$. It is well seen in Fig.~\ref{fig:traj}(a) that the $x$-component of the orientation vector $n_x$
experiences decaying oscillations around 0, while $n_y$
converges to 1. This indicates that at the beginning particles exhibit a kayaking, which then slowly evolves to a log-rolling motion
(see Fig.~\ref{fig:sketch}). Note that oscillations in the orientation of a  spheroid with $b/a=0.8$ decrease much slower than those for a spheroid of $b/a=0.5$.
The kayaking motion is responsible for the
oscillations in trajectories shown in Fig.~\ref{fig:traj}(b). We see that for a spheroid of $b/a=0.8$ the migration to
the equilibrium position is faster than for that of $b/a=0.5$, although the particle trajectory is
much less affected by the kayaking motion. Another important observation is that the equilibrium positions for all spheroids are very close, pointing strongly that
they are defined by the
equatorial radius $a$. This result is consistent with reported experimental data~\cite{hur2011, masaeli2012}.
 To validate this finding, below we compute
$z_{eq}$ for spheroids of different sizes and aspect ratios.

\begin{figure}
\vspace{-0.4cm}
\centering
\includegraphics[width=1.0\columnwidth]{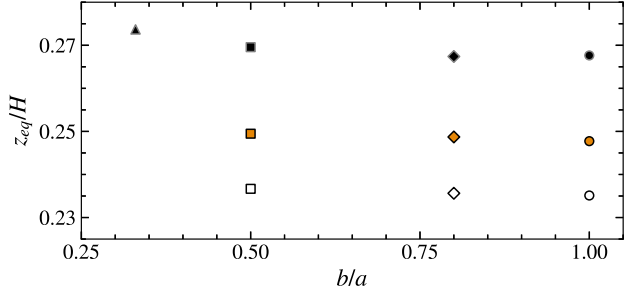}
\vspace{-1.2cm}
\caption{Equilibrium positions of spheroids with fixed
	$a/H=0.075$, $0.1$, $0.15$ (open,colored, black symbols, respectively)
	vs. their aspect ratio.}
\label{fig:zeq_ba}
\end{figure}

Let us now fix several $a$ and simulate $z_{eq}$ as a function of $b/a$.
The results for a lower equilibrium position are plotted in Fig.~\ref{fig:zeq_ba}. As expected,
$z_{eq}$ strongly depends on $a$, but is practically independent of
the aspect ratio of spheroids. We stress that equilibrium positions are nearly independent of $\Re$ in the range from 11 to 44 used here. The same conclusion has been earlier made for spherical particles~\cite{asmolov2018}.

Based on these observations, one can speculate that the lift coefficient at any, not only equilibrium,
$z$ is controlled by the equatorial radius.
If so, we can suppose that the lift force on oblate spheroids of equatorial radius $a$  represents a product of Eq.~\eqref{lift} for a sphere of the same radius and the correction $f$ that depends on the aspect ratio

\begin{equation} F_l=\rho a^4 G_m^2c_l(z/H,a/H,\Re)f(b/a) .
\label{factorization}
\end{equation}
This ansatz constitutes nothing more than an assumption, made to provide the lift force that depends on $z$ only through the lift coefficient $c_l$.

\begin{figure}[tbp]
\centering
\vspace{-0.5cm}
\includegraphics[width=1.0\columnwidth]{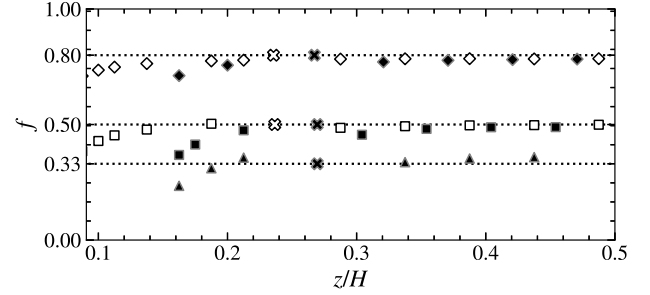}
\vspace{-1.1cm}
\caption{Ratio of the lift forces for spheroids and spheres of the same $a$ computed at $\Re=22$ using $a/H=0.075$ (open symbols) and $0.15$ (black symbols). The aspect ratio $b/a$ of spheroids in simulations is set to be equal to $0.33$ (triangles), $0.5$ (squares), and $0.8$
	(diamonds). Equilibrium positions of spheroids are marked by open ($a/H=0.075$) and black ($a/H=0.15$) crosses. Dotted lines show $f = b/a$.}
\label{fig:ratios}
\end{figure}

To verify Eq.~\eqref{factorization}, we fix the $z$ position of a spheroid that exhibits a stable log-rolling motion  but is free to also translate
in two other directions, and measure the lift force. If the form of ansatz~\eqref{factorization} is correct, the ratio
 $F_l/\left(\rho a^4 G_m^2c_l\right)$ would be equal to
$f(b/a)$. In Fig.~\ref{fig:ratios} we plot this ratio as a function of the particle
position and conclude that for a given $b/a$ it is nearly constant. Moreover, we see that $f \simeq b/a$. Note that results displayed in Fig.~\ref{fig:ratios} correspond to $\Re = 22$, but these conclusions have been verified for $\Re=11$ and 44 (not shown). Therefore, one can rewrite Eq.\eqref{factorization} as
\begin{equation}
F_l=\rho a^3 b G_m^2 c_l(z/H,a/H,\Re).
\label{scaling} \end{equation}

\begin{figure}
\centering
\vspace{-0.4cm}
\includegraphics[width=1.\columnwidth]{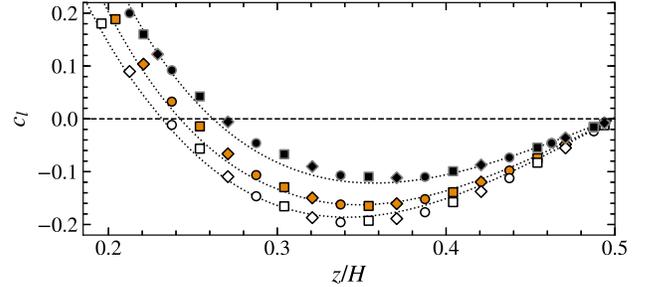}
\vspace{-1.2cm}
\caption{Lift coefficients, Eq.~(\ref{scaling}), in the channel central portion of the channel at
$\Re=22$ computed using $a/H=0.075$, $0.1$, and $0.15$ (open, colored, black symbols,	 respectively).  The aspect ratio $b/a$ is equal to $0.5$ (squares), $0.8$ (diamonds) and $1$ (circles).Dotted curves show calculations from Eq.(\ref{cli}) using Eqs.(\ref{cl_0})-(\ref{cl_2}) for $c_{li}$.}
\label{fig:scaling}
\end{figure}

Eq.\eqref{scaling} allows one to obtain $c_l$ from the simulation data simply by computing the ratio $F_l/(\rho a^3 b G_m^2)$, which is expected to depend on $a/H$, but not on the spheroid aspect ratio. We now calculate $c_l$ for a sphere and spheroids of two aspect ratios (0.5 and 0.8, as before) using several values of $a/H$. The
simulation results for the central region of the channel, $0.2 \leq z/H \leq 0.5$ are given in Fig.~\ref{fig:scaling}, which fully confirms that at
fixed $a/H$ the lift coefficient indeed does not depend on $b/a$. Using these simulation results in Appendix~\ref{ap:fit} we propose fitting expressions for the lift coefficient. Calculations from Eq.(\ref{cli}) using $c_{li}$ given by Eqs.(\ref{cl_0})-(\ref{cl_2}) are also included in Fig.~\ref{fig:scaling}, and we see that they fit well the simulation dataThe overall conclusion from this plot is that our scaling Eq.\eqref{scaling} adequately describes the lift force in the
central region of the channel.

\begin{figure}
\vspace{-0.5cm}
\centering
\includegraphics[width=1.0\columnwidth]{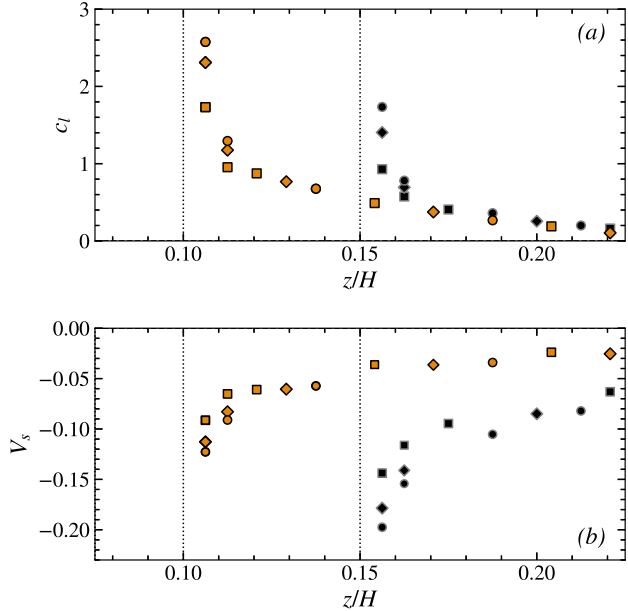}
\vspace{-1.2cm}
\caption{(a) Lift coefficient and (b) particle slip velocity near the wall for spheroids
	of $b/a=0.5$ (squares), $0.8$ (diamonds) and spheres (circles). Colored and black symbols correspond to
	$a/H=0.1$ and $0.15$. Dotted lines indicate a contact with the wall.}
\label{fig:scaling2}
\end{figure}

 However, as seen in Fig.~\ref{fig:scaling2}(a), Eq.\eqref{scaling} becomes inaccurate very close to the wall, namely, when $z-a\leq 0.2a$. At such small distances between spheroids and the wall the lift coefficient is no longer independent on the aspect ratio, and we see that $c_l$ augments with $b/a$. An explanation for the smaller $c_l$ for the spheroids compared to the sphere can be obtained if we invoke their hydrodynamic interactions with the wall that depends on both $a$ and $b$~\cite{vinogradova1996}.
This is illustrated and confirmed in
Fig.~\ref{fig:scaling2}(b), where the data for the particle slip velocity near the wall are presented. It is well seen that close to the wall $V_s$
is finite and varies with both $a$ and $b/a$.  More oblate particles have a smaller slip
velocity and, therefore, experience a smaller lift force.

\begin{figure}
\centering
\vspace{-0.4cm}
\includegraphics[width=1.0\columnwidth]{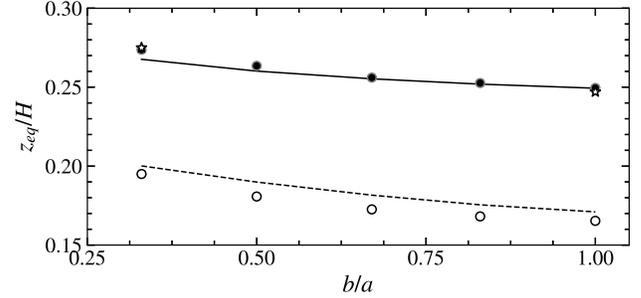}
\vspace{-1.2cm}
\caption{Equilibrium positions for spheroids of the same volume (equivalent  to that of a
	sphere of $a/H=0.1$) vs. the aspect ratio obtained in simulations at $\Re=22$ (circles). Black circles indicate
	neutrally buoyant spheroids, open circles show results for non-neutrally buoyant spheroids subject to an external force ($c_{ex}=-0.045$). Stars show the simulation data by  \citet{lashgari2017} obtained at $\Re=50$.
	Solid and dashed curves are calculations from Eqs.~(\ref{scaling}) and (\ref{zf_eq}). In both cases Eq.~(\ref{cli}) and Eqs.(\ref{cl_0})-(\ref{cl_2}) are used to calculate $c_l$ and $c_{li}$. }
\label{fig:zeq_vol}
\end{figure}

Additional insight into the problem can be gleaned by computing the equilibrium positions for particles of an equal volume $V$, but of various aspect ratio. This situation is relevant to separation experiments~\cite{hur2011,masaeli2012}. We now fix $a^{2}b$, so that it is equivalent of that for a sphere of $a/H = 0.1$,  and measure $z_{eq}$ at $\Re=22$ and different $b/a$. Simulation results for neutrally-buoyant oblate spheroids are included in Fig.~\ref{fig:zeq_vol}  (black symbols). It is seen that a decrease in $b/a$ has the effect of larger $z/H$, although rather insignificant. Since the lift
coefficient and $z_{eq}$ (where $c_l$ vanishes) are independent of $b$ as follows from Eq.~\eqref{scaling}, the weak variations of $z_{eq}$ with the aspect ratio are caused by the changes in values of $a$. Fig.~\ref{fig:zeq_vol} also includes the data obtained by Lashgari {\it et al.} \cite{lashgari2017} by means of the LBM simulations at $\Re=50$ (shown by stars). We see that their results agree well with our simulations thus confirming that at moderate $\Re$ the equilibrium positions do not depend on its values.
Finally, we note that calculations from Eq.(\ref{scaling}) using Eq.~(\ref{cli}) for $c_ l$ and Eqs.(\ref{cl_0})-(\ref{cl_2}) for $c_{li}$ fit the simulation data very well (solid curve). 

These simulations are compared with analogous, made with the same parameters, but in which $\rho_p \neq \rho$ and an external force is incorporated. The equilibrium positions of such non-neutrally buoyant spheroids have been found from
\begin{equation}
c_{l}(z_{eq}/H,a/H,\Re )=-\dfrac{c_{ex}H}{a},
\label{zf_eq}
\end{equation}
obtained by using Eq.\eqref{eq:balance} together with Eq.~(\ref{scaling}), where the dimensionless parameter $c_{ex}$ that characterizes the relative value of
the external force is given by
\begin{equation}\label{eq:cex}
  c_{ex}=\frac{4\pi f_{ex}}{3\rho  G_{m}^{2}H}.
\end{equation}
Since Eq.(\ref{zf_eq}) does not include $b$, at constant $f_{ex}$ the equilibrium positions for particles of the same $a$ coincide.
Simulations made with $c_{ex}=-0.045$
are included in Fig.~\ref{fig:zeq_vol}  (open symbols) and show that $z_{eq}/H$ are shifted towards the bottom wall compared to neutrally buoyant spheroids. Note that with our parameters Eq.~(\ref{zf_eq}) has only one root, so that  the upper equilibrium position cannot be attained.
Also included in Fig.~\ref{fig:zeq_vol} are the calculations from Eq.~(\ref{zf_eq}), where $c_l$ is obtained using Eq.(\ref{cli}) with $c_{li}$ given by (\ref{cl_0})-(\ref{cl_2}) (dashed curve). We see that they are in a good agreement with the simulation results.

\section{Conclusion}\label{sec:conclusion}

We have presented lattice Boltzmann simulation data on the inertial migration of oblate spheroids in the channel flow with
moderate Reynolds numbers. Our results show that spheroids focus to equilibrium positions, which depend only on their equatorial radius $a$, but not on the polar radius $b$. We invoke this simulation result to derive a scaling expression for a lift force, Eq.(\ref{scaling}). In this expression, the lift force is proportional to $a^3 b$, but the lift coefficient, $c_l$, is the same as for a sphere of radius $a$. We have also proposed fitting expressions allowing one to easily calculate $c_l$. Our scaling theory is shown to be valid throughout the channel, except  very narrow regions near a wall. Thus, it can be employed to  predict, with high accuracy, the
equilibrium positions of spheroids in the channel. These,  in turn, could be used  to develop inertial microfluidic methods for a
shape-based separation.

We recall, that in our work we have limited ourselves by oblate spheroids of $b/a \geq 0.3$ and used $\Re \leq 44$ only,
but one cannot exclude that at lower aspect ratios and/or larger Reynolds numbers the equilibrium positions would depend on both radii of particles. It would be of considerable interest to explore the validity of Eq.~(\ref{scaling}) using other flow and oblate spheroid parameters. Another fruitful direction could be an investigation of prolate particles to develop an analogue of
Eq.~(\ref{scaling}).


\begin{acknowledgments}
This research was partly supported by the Russian Foundation for Basic Research
(grant 18- 01-00729), by the Ministry of Science and Higher Education
of the Russian Federation and by the German Research Foundation (research
unit FOR2688, project HA4382/8-1).
\end{acknowledgments}

\section*{Data availability statement}
The data that support the findings of this study are available
within the article.

\appendix
\section{Fitting expressions for  the lift coefficients and particle velocity}\label{ap:fit}

The lift coefficient for the sphere is given by Eq.(\ref{cli}) and depends on the coefficients $c_{li}$ and on the slip velocity $V_s$. The later is defined by Eq.(\ref{Vslip}) and depends on the particle velocity $V_p^x$. Consequently, to apply Eq.\eqref{scaling} for oblate spheroids we have to determine $c_{li}$ and $V_p^x$ for a sphere. In this Appendix we propose some useful fitting expressions for these functions.

We propose a modification of expressions for  $c_{li}$ reported by  \citet{asmolov2018}
\begin{equation}
c_{l0}=\beta_0(a/H)c_{l0}^{VC}(z/H),
\label{cl_0}
\end{equation}
\begin{equation}
c_{l1}=4(1-2z/H)\beta_1(a/H)c_{l1}^{CM}(z/a),
\label{cl_1}
\end{equation}
\begin{equation}
c_{l2}=c_{l2}^{CM}(z/a),
\label{cl_2}
\end{equation}
where correction factors $\beta_0$ and $\beta_1$ allow for fiting the data for bigger particles then before. By fitting our simulation data for several $a/H$ (see Sec.~\ref{sec:results}) we obtain
\begin{equation}
\begin{array}{ll}
\beta_0=1+3.32(a/H)-26.45(a/H)^2,\\
\beta_1=1-8.39(a/H)+19.65(a/H)^2.
\end{array}
\end{equation}

The coefficient $c_{l0}^{VC}$ in Eq.\eqref{cl_0} represents the analytical solution by \citet{VasseurCox1976} for pointlike spherical particles in the low $\Re$ channel flow, which can be well fitted by
\begin{equation}
c_{l0}^{VC}=2.25\left( z/H-0.5\right) -23.4\left( z/H-0.5\right) ^{3}.
\label{cl0}
\end{equation}
 as suggested by \citet{feuillebois2004}.
The coefficients $c_{l1}^{CM}$ and $c_{l2}^{CM}$ in Eqs.\eqref{cl_1} and \eqref{cl_2} are these obtained by Cherukat and McLaughlin\cite{Cheruk:Mclau94,Cheruk:Mclau95} for finite-size particles in a near-wall shear flow
\begin{equation}
c_{l1}^{CM}=-3.2415\zeta -2.6729-0.8373\zeta ^{-1}+0.4683\zeta ^{-2},
\label{cl1CM}
\end{equation}%
\begin{equation}
c_{l2}^{CM}=1.8065+0.89934\zeta ^{-1}-1.961\zeta ^{-2}+1.02161\zeta ^{-3},
\label{cl2CM}
\end{equation}
where $\zeta = z/a$.

At relatively large distances from the wall, the particle slip velocity (and, hence, $V_p^x$)  is independent of $b$ and,unlike wall-bounded shear flow, remains finite (see Fig.~\ref{fig:scaling2}).  Therefore, for this central region one can use  the approximation for the sphere of radius $a$. 
The velocity $V_p^x$ can be presented as a sum of the solution for a wall-bounded linear shear flow~\cite{reschiglian2000,wakiya1967}
 \begin{equation}
V^x_{l}=U\left( z\right) h(z/a),
  \label{gold}
\end{equation}%
with 
\begin{equation}
h=\frac{200.9\xi-\left( 115.7\xi+721\right) \zeta^{-1} -781.1}{%
-27.25\xi^{2}+398.4\xi-1182}\quad \mathrm{at}\quad \zeta<3,
  \label{rech}
\end{equation}%
\begin{equation}
h=\frac{1-\frac{5}{4}\zeta^{-3}+\frac{5}{4}\zeta^{-5}-\frac{23}{48}\zeta^{-7}-%
\frac{1375}{1024}\zeta^{-8}}{1-\frac{15}{16}\zeta^{-3}+\zeta^{-5}-\frac{3}{8}%
\zeta^{-7}-\frac{4565}{4096}\zeta^{-8}}\quad \mathrm{at}\quad\zeta\geq 3,
  \label{wak}
\end{equation}
where $\xi=\log (\zeta-1),$
and the Faxen correction due to the parabolic flow profile:
\begin{equation}
V_p^x=V_{l}^x-4/3(a/H)^2.
\label{Vp}
\end{equation}

\end{document}